\mathchardef\tildechar="707E
\def\dntilstyle#1{{\raisebox{-0.85\height}[.1\height][0pt]{$#1\tildechar$}}}

\def\wavefillstyle#1{\cleaders%
\hbox{$#1\mkern-2.5mu{\dntilstyle{#1}}\mkern-2.5mu$}\hfill}

\def\overwavestyle#1#2{\vbox{\ialign{##\crcr\noalign{\kern.2ex}
      \wavefillstyle{#1}\crcr\noalign{\kern.2ex\nointerlineskip}
      $\hfil#1#2\hfil$\crcr}}}
\def\underwavestyle#1#2{\vtop{\ialign{##\crcr
      $\hfil#1#2\hfil$\crcr\noalign{\kern.2ex\nointerlineskip}
      \wavefillstyle#1\crcr\noalign{\kern.2ex}}}}

\def\overwave#1{\mathchoice{\overwavestyle{\displaystyle}{#1}}%
{\overwavestyle{\textstyle}{#1}}{\overwavestyle{\scriptstyle}{#1}}%
{\overwavestyle{\scriptscriptstyle}{#1}}}

\documentclass[12pt]{article}
\usepackage{amssymb,amsmath}

\let\ox=\otimes
\let\iso=\cong

\newcommand\C{\mathbb{C}}
\newcommand\R{\mathrm{I\!R}}
\renewcommand\H{\mathbb{H}}
\newcommand\h{{\scriptstyle\mathbb{H}}}
\newcommand\Sph{\mathbb{S}}

\newcommand{\mx}[6]{{\left#1\begin{array}{rr}#2&#3\\#4&#5\end{array}\right#6}}
\newcommand{\smx}[3]{{\left#1\begin{smallmatrix}#2\end{smallmatrix}\right#3}}
\newcommand{\Mx}[3]{\left#1\begin{array}{cccc}#2\end{array}\right#3}

\newcommand{\Id}{\mathfrak{1}}
\newcommand{\I}{\mathfrak{i}}
\newcommand{\J}{\mathfrak{j}}
\newcommand{\K}{\mathfrak{k}}
\newcommand{\q}{\mathbin\mathfrak{q}} %mathbin for better spacing
\newcommand{\sx}{\sigma_{\!x}}
\newcommand{\sy}{\sigma_{\!y}}
\newcommand{\sz}{\sigma_{\!z}}
\newcommand{\gm}[1]{\gamma_{#1}}

\newcommand{\Op}[1]{\boldsymbol{#1}}
\newcommand{\Er}{I\!\!E}
\newcommand{\Imat}{\mathrm{1\!\!I}}

\newcommand{\EQ}[1]{\mbox{$#1$}}

\newcommand{\ket}[1]{|{#1}\rangle}
\newcommand{\bra}[1]{\langle{#1}|}

\newcommand{\brkt}[2]{\langle #1 \mid #2 \rangle}
\newcommand{\brct}[2]{\langle #1 \,, #2 \rangle}
\newcommand\0{\ket{0}}
\newcommand\1{\ket{1}}
\newcommand\qE{{\mathbf{0}}}
\newcommand\qI{{\mathbf{\dot{0}}}}
\newcommand\qJ{{\mathbf{1}}}
\newcommand\qK{{\mathbf{\dot{1}}}}
\newcommand\qPh{\vphantom{\dot{\qI}}}

\newcommand\QE{{\mathbf{0_0}}}
\newcommand\QI{{\mathbf{0_1}}}
\newcommand\QJ{{\mathbf{1_0}}}
\newcommand\QK{{\mathbf{1_1}}}

\newcommand{\vc}[1]{{\mathbf{#1}}}
\newcommand{\cc}[1]{\overline{#1}\vphantom{#1}}
\def\hc{{\dagger}}
\newcommand{\qc}[1]{\overwave{#1}}

\let\injmap=\hookrightarrow

\newcommand{\proj}{\mathrel{{\triangleright}\!{\to}}}
\newcommand{\Proj}{\mathrel{{\triangleright}\!{\longrightarrow}}}
\newcommand{\isoeq}{\mathbin{\rightleftharpoons}}
\let\swap=\leftrightarrow
\newcommand{\myto}{\mathrel{\rightsquigarrow}}

\newcommand{\eq}[1]{Eq.~(\ref{#1})}
\def\DoListComSem#1#2<<#3,#4>>{#1{#3}\TailSem{#1}{#2}{#4}}
\def\TailSem#1#2#3{\ifx#3;\relax\else{#2}\DoListComSem{#1}{#2}<<#3>>\fi}
\def\DoList#1#2#3{{\DoListComSem#1#2<<#3,;>>}}
\newcommand{\eqs}[1]{Eqs.~(\DoList{\ref}{,~}{#1})}

\newcommand{\pd}[1]{\frac{\partial}{\partial{#1}}}
\def\+{\phantom{-}}

\def\Sum{\sum\limits}
\newcommand\D[1]{{\tt #1D}}

\begin{document}

\title{Error Correction with Euclidean Qubits}

\author{{\em Alexander Yu.\ Vlasov}\thanks{%
{\tt E-mail: Alexander.Vlasov@pobox.spbu.ru}}}
\date{November 1999}

\maketitle

\begin{abstract}
 In classical case there is simplest method of error correction with using
three equal bits instead of one. In the paper is shown, how the scheme fails
for quantum error correction with complex vector spaces of usual quantum
mechanics, but works in real and quaternionic cases. It is discussed also,
how to implement the three qubits scheme with using encoding of quaternionic
qubit by Majorana spinor. Necessary concepts and formulae from area of
quantum error corrections are closely introduced and proved.
\end{abstract}

\section{Introduction}

A simple question, why quantum error correction code for one qubit needs
for 5, or 7, or 9 \ldots qubits instead of 3 in classical case, is concerned
with some rather deep topics. In the article is shown that difference between
Hilbert and Euclidean spaces is also matter here, for example, three
{\em Euclidean} qubits would be enough and it is discussed below.

Together with obvious example with real vector spaces, Euclidean case also
related with quaternionic representation of qubits and quantum gates.
Noncommutative algebra of quaternions is richer than complex numbers
and can represent some nonstandard view on a qubit.

To show, that the view has some relation with physical reality the
{\em quaternionic} qubit may be considered as some subsystem of four component
Dirac spinoral wave function (end of Sec.\ref{Sec:Maj}), but the relativistic
example should not be considered as only way of interpretation of the
modified qubit model.

The Sec.\ref{Sec:RC} devoted to simple case of error correction
for {\em real} vector space. A closed introduction to topics of quantum
error correction necessary for questions under consideration is given
in Sec.\ref{Sec:CEC}. Quaternionic case is introduced in Sec.\ref{Sec:QQ}
and example with Majorana spinors is described in Sec.\ref{Sec:Maj}.

\subsection*{Standard definitions}

The Pauli matrices are:
$$\sx = \mx (0110),\ \sy = \mx (0{-i}i0),\ \sz = \mx (100{-1})$$

$\R$ --- real numbers, $\C$ --- complex numbers.

$\C^n_\star = \C^n\!{-}\{\mathbf0\}$ --- vector space without origin.

$\Sph^n$ --- unit sphere in $\R^{n+1}$,
{\em i.e.} $\Sph^2$ --- usual sphere and $\Sph^1$ --- circle.

$\C{P}^n = \C^{n+1}_\star\!/\C^{\,}_\star$,
$\R{P}^n = \R^{n+1}_\star\!/\R^{\,}_\star$ --- complex and real projective spaces.

{\em Maps\/}: $M \proj N$ --- projection, $N \injmap M$ --- injection.

$A \iso B$ --- isomorphism of groups, algebras (also equivalence of two
               topological spaces).

$a \isoeq b$ means: $a \in A$, $b \in B$, $A \iso B$,
$A \stackrel{\iota}{\to} B$,  $\iota(a) = b$

$a \simeq b$ --- equivalence relation for elements ( used in definition
of quotient spaces like $SU(2)/U(1)$ ).

\medskip

All other necessary definitions ($\H$, $\gm{i}$, {\em etc.}) are given
below in main text of the paper.

\section{$SO(2)$ (or $U(1)$) error correction}\label{Sec:RC}

A trivial classical error correction method for one bit uses simplest 3 bit
encoding: $0 \to 000$, $1 \to 111$. Let us consider, as toy model, qubits with
{\em real} coefficients: $a\0 + b\1$, $(a,b\in\R)$. The $\R$-qubit is described
by some point on \D2 plane. Let us suggest now, that error --- is rotation of
the plane for one of the $\R$-qubits:
\begin{equation}
\0 \to \alpha \, \0 + \beta \, \1; \ \1 \to -\beta \, \0 + \alpha \, \1;
\quad \alpha = \cos \, \theta; \ \beta = \sin \, \theta
\label{SO}
\end{equation}

Analog of the classical scheme for $\R$-qubits is code:
\begin{equation}
 \0 \to \ket{000},\ \1 \to \ket{111};
 \quad a\0 + b\1 \to a\ket{000} + b\ket{111}
 \label{ooo}
\end{equation}
then any 1-qubit error \eq{SO} can be corrected. It is enough to
append some auxiliary {\em ancilla} qubits and to apply special
transformation to ``transfer'' an error on the extra qubits. In the
example under consideration we can add two ancilla qubits and use
orthogonal transformation\footnote{Only nontrivial
transformations of 8 basis vectors (between 32) are shown.} :

\begin{equation}
\begin{array}{rcr|rcr}
\ket{000}\ket{00} & \to & \ket{000}\ket{00} \ &
\ket{111}\ket{00} & \to & \ket{111}\ket{00} \\
\ket{100}\ket{00} & \to & \ket{000}\ket{10} \ &
\ket{011}\ket{00} & \to & -\ket{111}\ket{10} \\
\ket{010}\ket{00} & \to & \ket{000}\ket{01} \ &
\ket{101}\ket{00} & \to & -\ket{111}\ket{01} \\
\ket{001}\ket{00} & \to & \ket{000}\ket{11} \ &
\ket{110}\ket{00} & \to & -\ket{111}\ket{11}
\label{Rcorr}
\end{array}
\end{equation}

Let us consider error in first qubit as an example:
$$ a\,\ket{000} + b\,\ket{111} \myto
   a\, (\alpha_1 \ket{000} + \beta_1 \ket{100}) +
   b\, (-\beta_1 \ket{011} + \alpha_1 \ket{111}) \myto
$$
After appending ancilla $\ket{00}$:
$$ \myto
   a\, (\alpha_1 \ket{000}\ket{00} + \beta_1 \ket{100}\ket{00}) +
   b\, (-\beta_1 \ket{011}\ket{00} + \alpha_1 \ket{111}\ket{00}) \myto
$$
And after application of operator \eq{Rcorr} :
\begin{eqnarray*}
&\myto&  a\, (\alpha_1 \ket{000}\ket{00} + \beta_1 \ket{000}\ket{10}) +
    b\, (\beta_1 \ket{111}\ket{10} + \alpha_1 \ket{111}\ket{00}) \\
&=& (a\,\ket{000} + b\,\ket{111}) \, (\alpha_1\ket{00} + \beta_1\ket{10})
\end{eqnarray*}

\smallskip

The scheme works also for usual ($\C$-)qubit with complex coefficients $a,b$
and only important condition is \eq{SO} with real $\alpha, \beta$ , {\em i.e.}
$SO(2)$ group of errors. Similar 3-qubit scheme was used for experimental
{\em phase} error correction \cite{ExpEC}. It is possible because
$U(1) \iso SO(2)$ and so phase, $U(1)$, error can be considered as $SO(2)$
error in other basis\footnote{For paper \cite{ExpEC} it is
$\ket{+} = (\0+\1)/\sqrt{2}$ and $i\,\ket{-} = i\,(\0-\1)/\sqrt{2}$ .}.

\section{Quantum (complex) error correction}{\label{Sec:CEC}}

The method above does not work for more general set of errors. Error matrix in
\eq{SO} could be written as:
$$ \Mx({\alpha&\beta\\-\beta&\alpha}); \quad
\alpha,\beta \in \R; \ \alpha^2+\beta^2=1 $$
and general 1-qubit error can be expressed as element of $SU(2)$ group:
\begin{equation}
\Mx({\alpha&\beta\\-\cc\beta&\cc\alpha}); \quad
\alpha,\beta \in \C; \ |\alpha|^2+|\beta|^2=1
\label{SU}
\end{equation}
If we use code like \eq{ooo} with complex $a,b$, then we can correct error
\eq{SU} if $\alpha$ and $\beta$ are real, but if, for example, any qubit
suffers `$\Op{E}_\pi$' error with $\alpha = i$, $\beta = 0$; $\0 \to i\0$,
$\1 \to -i\1$ then $a\,\ket{000} + b\,\ket{111}$ is transformed as
$a\,i\ket{000} - b\,i\ket{111}$ and it could not be distinguished from
case without error, but with initial coefficients $a' = i\,a$, $b' = -i\,b$.
Because only pairs with same phase multiplier are physically equal,
$(a,b) \simeq (\phi\,a,\phi\,b)$, the $(a',b')$ correspond to other state
if $a \neq 0$ and $b \neq 0$.

\smallskip

Generally, quantum error correction suggests also entanglement of qubit
with environment \cite{ShEC}.

\medskip

The extension of error correction method discussed in beginning of
Sec.\ref{Sec:RC} is following. Let the $\ket{w_l}$ is codeword for $\ket{l}$,
$\Op{E}_p$ is some error operator and $\ket{0_A}$, $\ket{A_p}$ are initial and
final states of ancillas. Then unitary error correction operator
$\Op{U}_{ec}$ acts as:
\begin{equation}
 \Op{U}_{ec} \bigl( (\Op{E}_p \ket{w_l}) \ket{0_A} \bigr)
 = \ket{w_l} \ket{A_p}
\label{GenEC}
\end{equation}

To show, that the operator $\Op{U}_{ec}$ corrects a linear combination of
errors, let us consider a state $ \ket{W} = \sum_l c_l \ket{w_l}$ and error
operator $\Op{\Er} = \sum_p e_p \Op{E}_p$:
\begin{equation}
 \begin{array}{rclcl}
 \Op{U\!}_{ec}\bigl((\Op{\Er}\ket{W})\ket{0_A}\bigr)
  &=& \Sum_{p,l} c_l e_p \bigl( (\Op{U\!}_{ec} \Op{E}_p \ket{w_l} )\ket{0_A}\bigr)
      &=& \Sum_{p,l} c_l e_p\ket{w_l} \ket{A_p} \rlap{\ =} \\ \\
  &=& \Sum_l c_l \ket{w_l} \, \Sum_p e_p \ket{A_p}
      &=& \ket{W} \, \Sum_p e_p \ket{A_p}
 \end{array}
\label{UEr}
\end{equation}

The same expression \eq{UEr} is valid for entanglement with environment if
to consider instead of complex numbers $e_p$ operators $\Op{e_p}$ those act
on environment term in product $\ket{\mbox{Env}}\ox\ket{W}$, {\em i.e.}
$\Op{\Er} = \sum_p \Op{e}_p\ox\Op{E}_p$.

To write instead of \eq{GenEC} two standard (see \cite{BDSW,EO,KLEC} )
conditions, let us note, the unitary operator
$\Op{U}_{ec}$ does not change scalar products of any two vectors, {\em i.e.}:
\begin{equation}
 \bra{w_{l_1}} \Op{E}_{p_1}^\hc \Op{E}_{p_2}^{\,} \ket{w_{l_2}} =
 \brkt{w_{l_1}}{w_{l_2}} \, \brkt{A_{p_1}}{A_{p_2}}
\label{GenECn}
\end{equation}
where expression includes also case with $\Op{E}$ is no-error, {\em i.e.}
identity operator and $\Op{E}_p^\hc$ is Hermitian conjugation,
$\brct{\Op{E}_p w}{u} = \brct{w}{\Op{E}_p^\hc u}$. Two different
cases: $l_1 \neq l_2$ and $l_1 = l_2$ produce two sets of equations:
\begin{eqnarray}
 \bra{w_{l_1}} \Op{E}^\hc_{p_1} \Op{E}_{p_2}^{\,} \ket{w_{l_2}} & = & 0
 \qquad (l_1 \neq l_2)
 \label{GenECone}  \\
 \bra{w_{m}} \Op{E}^\hc_{p_1} \Op{E}_{p_2}^{\,} \ket{w_{m}} & = &
 \bra{w_{n}} \Op{E}^\hc_{p_1} \Op{E}_{p_2}^{\,} \ket{w_{n}}
 \label{GenECtwo}
\end{eqnarray}

\smallskip

With using \eqs{GenECone,GenECtwo} it is possible to show, why
{\em non-entangled} code like \eq{ooo} does not work with $SU(2)$ error.
For one qubit {\em phase} error ($\Op{E}_\pi$) discussed earlier
$\bra{0}\Op{E}_\pi\ket{0} = i$, $\bra{1}\Op{E}_\pi\ket{1} = -i$ does not compatible
with \eq{GenECtwo} and the wrong multiplier breaks simple `product' code
\eq{ooo}.

On the other hand, the same \eq{GenECtwo} shows that for {\em effective}\footnote{
{\em I.e} different $\Op{E}_p$ map same vector in orthogonal linear spaces}
qubit error like \eq{SO} we have $\bra{l}\Op{E}_p\ket{l} = 0$ and code
would work. It should be mentioned, the discussed {\em effectiveness}
condition is sufficient, but not necessary for error correction code.

\section{Quaternionic qubits}{\label{Sec:QQ}}

\subsection{Preliminaries}

The {\em quaternions}, $\H$ --- are \D4 (real) algebra with basis $\I, \J, \K$,
and $\Id$ (unit), $\I^2=\J^2=\K^2 = -\Id$, $\I\J=-\J\I=\K$, $\J\K=-\K\J=\I$,
$\K\I=-\I\K=\J$. It is algebra with multiplicative norm, like complex
numbers, {\em i.e.} for $\vc{q} = q_0\Id + q_1\I + q_2\J + q_3\K$ norm is
Euclidean length of $\vc{q}$: $|\vc{q}|^2 = q_0^2 + q_1^2 + q_2^2 + q_3^2$
and $|\vc{q}|\,|\vc{h}| = |\vc{q\,h}|$.

Quaternionic conjugation is introduced as
$\qc{\vc q}\equiv q_0\Id - q_1\I - q_2\J - q_3\K$, with properties:
$\vc{q}\,\qc{\vc q} = |\vc{q}|^2$, $\qc{\vc{q\,u}}=\qc{\vc{u}}\,\qc{\vc{q}}$,
$\vc{q}^{-1} = \qc{\vc{q}}/|\vc{q}|^2$.
Quaternions may be used for representation of \D3 rotations; if $\vc{v}$ is
pure imaginary quaternion, {\em i.e.} $v_0=0$ (or $\qc{\vc{v}} = - \vc{v}$)
any rotation of vector $\vc{v}$ can be represented as:
\begin{equation}
 \vc{v}' = \vc{q}\,\vc{v}\,\vc{q}^{-1}; \quad \mbox{or simply }
 \ \vc{v}' = \vc{q}\,\vc{v}\,\qc{\vc q},\ |\vc{q}| = 1
\label{QRot}
\end{equation}

\smallskip

Qubit $a\0+b\1$ ($a=a_x+i\,a_y$, $b=b_x+i\,b_y$) can be considered as element
of \D2 complex or \D4 real vector space and can be expressed as quaternion
\EQ{\q = a_x \Id + a_y \I + b_x \J + b_y \K} or simply $\q = a + b\,\J$.
Here `usual' complex $i$ is equivalent with {\em left} multiplication on $\I$,
for example: \EQ{\exp(\varphi\I){\cdot}\q = e^{i\varphi}a + e^{i\varphi}b \J}.

With the notation, physically equivalent states could be described as:
\begin{equation}
\q \simeq e^{\varphi\I}\q
\label{qfaz}
\end{equation}
Where $\varphi$ is a real number and $|\q| = 1$.
The action of $SU(2)$ group in \eq{SU} is expressed via {\em right}
multiplication. To show it, let us consider $\vc{u} = c + d\J$,
$\qc{\vc{u}} = \qc{c + d\J} = \cc{c} - d\J$ and so:
$$
\begin{array}{rclcl}
\q \vc{u} &=& (a + b\J)\,(c + d\J) &=& ac - b\cc{d} + (ad + b\cc{c})\J \\
\q \qc{\vc{u}} &=& (a + b\J)\,(\cc{c} - d\J) &=&
 a\cc{c} + b\cc{d} + (-ad + bc)\J
\end{array}
$$
or
\begin{equation}
\q \qc{\vc{u}} \isoeq \Mx({\cc{c}&\cc{d}\\-d&c}) \Mx({a\\b})
\quad (|\vc{u}| = |c|^2 + |d|^2 = 1)
\label{qSU}
\end{equation}
It is equivalent to \eq{SU} with $\alpha = \cc{c}$, $\beta = \cc{d}$.

Action of usual Pauli matrices in the notation is expressed as:
$$ \sx(\q) = \I \q \I,\ \sy(\q) = \I \q \J,\ \sz(\q) = \I \q \K$$
and arbitrary complex matrix can be expressed via:
\begin{equation}
 M(\q) = \q \vc{u} + \I \q \vc{w}
\label{CHisoM}
\end{equation}
It corresponds to isomorphism of algebra $\C\ox\H$ with algebra
of all $2\times2$ complex matrices.

\subsection{Qubit and Hopf fibration}

Here is only introduced {\em quaternionic notation} for usual ($\C^2$)
qubit. A quaternionic qubit will be discussed later. Now let us use
the notation for some simplification of description of usual qubit.

The qubit is example of mathematical object known as {\em Hopf fibration}:
$$
SU(2) \xrightarrow{U(1)\ } \Sph^2 \quad \mbox{or} \quad
\Sph^3 \xrightarrow{\Sph^1} \Sph^2
$$
Here normalized \D2 complex vectors correspond to quaternions with unit length
$|\q|=1$ and the subspace is isomorphic with $\Sph^3$ (sphere in \D4) or
with $SU(2)$ (see \eq{qSU}). The quotient of the space on equivalence relation
\eq{qfaz} is $SU(2)/U(1) \iso \Sph^2$ {\em i.e.} sphere in \D3.

A qubit $\ket\q$ maps to a sphere by projections $\Sph^3 \proj \Sph^2$
or $\C^2_\star \proj \Sph^2$:
\begin{equation}
 \q \Proj \vc{v} = \qc\q \,\I\,\q,\ |\q|=1;
 \quad (\mbox{or } \q \Proj \vc{v} = \q^{-1}\,\I\,\q,\ |\q| \ne 0)
 \label{sterprj}
\end{equation}
with properties:
\begin{eqnarray}
 e^{\varphi\I}\q & \Proj & \vc{v}' = \qc{e^{\varphi\I}\q}\,\I\,e^{\varphi\I}\q
= \qc\q e^{-\varphi\I} \, \I \, e^{\varphi\I} \q = \qc\q \, \I \, \q = \vc{v}
\label{fazprj} \\
 \q \qc{\vc{u}} & \Proj & \vc{v}' = \qc{\q \qc{\vc u}} \, \I \q \qc{\vc{u}}
= {\vc u} \, \qc\q \, \I \q \qc{\vc{u}} = \vc{u} \vc{v} \qc{\vc u}
\label{qSUprj}
\end{eqnarray}
The \eq{fazprj} shows that map \eq{sterprj} does not depend on phase
and meets \eq{qfaz}. The \eqs{QRot,qSUprj} show that unitary operation
\eq{qSU} corresponds to rotation of sphere $\Sph^2$.

The map \eq{sterprj} also can be considered as stereographic projection of
complex projective plane $\C{P} \iso \R^2{+}\{\infty\}$ to sphere $\Sph^2$, if
qubit $a\0+b\1$ is represented as element $a/b$ of $\C{P}$.

\smallskip

The description of qubit as Hopf fibration
%\begin{equation}
$ \Sph^3/\Sph^1 \to \Sph^2 $
%\end{equation}
here devoted to following problem.
We have two manifolds: $\Sph^3$ as space of normalized
wave vectors $|\psi|=1$ and $\Sph^2$ as physical space of states produced
by phase equivalence relation like \eq{qfaz}. The relation describes points
of some big circle on the sphere $\Sph^3$ and the circle maps to one point
on sphere $\Sph^2$. The sphere $\Sph^2$ forms {\em base} of Hopf fibration,
the big circle on $\Sph^3$ `over' a point of base is {\em fiber} and whole
$\Sph^3$ is {\em total space}.

The problem is: the physical space of states like Bloch sphere for spin
system corresponds to {\em base} and so is described by quite nonlinear way%
\footnote{``Nonlinear'' means, the space $\Sph^2$ does not accept some additive
structure.}. We introduce {\em linearity} of states in space $\C^2$ that
related with physical states by surjections:
$\C^2_\star \proj \Sph^3 \proj \Sph^2$.
The Hopf fibration let us manage with the last projection $\Sph^3 \proj \Sph^2$,
or $\Sph^3/\Sph^1 \iso \Sph^2$ as with a standard mathematical object.

The Hopf fibration is simplest example of
{\em nontrivial} fiber bundle {\em i.e.} the total space $\Sph^3$ does {\em not}
equivalent to direct product of base and fiber $\Sph^2\times\Sph^1$.
Physically it is related with following problem --- we may not consider
normalized wave vector for qubit $\ket\psi=a\0+b\1$, $|a|^2+|b|^2 = 1$ simply as
some pair $(\phi,s)$ with $\phi \in \Sph^1$ -- phase and $s \in \Sph^2$ --
phase-independent description of qubit state, for example point on Bloch sphere.

The other property of nontrivial bundle is absence of continuous inverse map
from points of base ($\Sph^2$) to fiber (big circle $\Sph^1$) over given point,
{\em i.e.} we cannot consider space of physically different states $\Sph^2$ as
some continuous subset of space $\Sph^3$ of complex 2-vectors with unit norm.

\smallskip

The property of qubit as Hopf fibration often makes rigorous mathematical
consideration of different constructions with qubit rather difficult.

\subsection{$\H$-qubits and $SU(2)$ error correction}

The {\em quaternionic qubit} ( $\h$-qubit ) is introduced here as \D1
quaternionic space {\em i.e.} \D4 real space with {\em Euclidean norm}
and action of group $SU(2)$ via right quaternionic multiplication.

$\H$-qubit can be considered as physical system with
state space isomorphic to $\Sph^3$ rather than $\Sph^2\iso\Sph^3/\Sph^1$.
The idea could be regarded as some allusion with {\em quaternionic
quantum mechanics\/}, but further in Sec.\ref{Sec:Maj} will be described an
application of the model to usual quantum mechanics by embedding quaternions
as \D4 real subspace in \D4 complex space of Dirac spinors.

\medskip

Let us denote basis of the space $\H$ as:
$$\Id\to\ket\qE,\ \I\to\ket\qI,\ \J\to\ket\qJ,\ \K\to\ket\qK$$
or
$$\Id\to\ket\QE,\ \I\to\ket\QI,\ \J\to\ket\QJ,\ \K\to\ket\QK$$

The two notations emphasize that $\h$-qubit extends 1-qubit system, but can
be included in 2-qubit space:
$$
\begin{array}{ccccc}
\C{P} & \injmap & \Sph^3 & \injmap & \C{P}^3 \\
\cap  &         & \cap   &         & \cap    \\
\C^2  & \iso    &  \H    & \subset & \C^4
\end{array}
$$

The $n$-$\h$-qubit space is introduced as \D{$4^n$} real space of tensor product:
$$\H^{{\ox}n} = \underbrace{\H\ox_\R \H\ox_\R \cdots \ox_\R \H}_n$$

\smallskip

Let us consider 3-$\h$-qubit error correction code $\{ \ket{\qE\qE\qE},
\ket{\qJ\qJ\qJ} \}$. The \mbox{1-$\h$-qubit} $SU(2)$ errors act via right
multiplication like in \eq{qSU}. The examples of error in first
$\h$-qubit are shown in next tables with two different notations:

\begin{equation}
\begin{array}{|l|l|l|l|}
\, \shortstack{code-\\word}
                 &\+\,\times\I     &\+\,\times\J     &\+\,\times\K  \\ \hline
\+\ket{\qE\qE\qE}&\+\ket{\qJ\qE\qE}&\+\ket{\qI\qE\qE}&\+\ket{\qK\qE\qE}\qPh\\
\+\ket{\qJ\qJ\qJ}& -\ket{\qE\qJ\qJ}& -\ket{\qK\qJ\qJ}&\+\ket{\qI\qJ\qJ}
\end{array}
\end{equation}
or
\begin{equation}
\begin{array}{|l|l|l|l|}
\, \mbox{codeword}&\+\,\times\I     &\+\,\times\J     &\+\,\times\K \\ \hline
\+\ket{\QE\QE\QE}&\+\ket{\QJ\QE\QE}&\+\ket{\QI\QE\QE}&\+\ket{\QK\QE\QE} \\
\+\ket{\QJ\QJ\QJ}& -\ket{\QE\QJ\QJ}& -\ket{\QK\QJ\QJ}&\+\ket{\QI\QJ\QJ}
\end{array}
\end{equation}

The $SU(2)$ errors acts {\em effectively} on one qubit, general error
correction conditions \eqs{GenECone,GenECtwo} are satisfied (in Euclidean
norm) for the 3-$\h$-qubit code and after appending few ancilla $\h$-qubits
any such errors can be corrected via orthogonal error correction operator.

\section{Example with Dirac, Majorana spinors}\label{Sec:Maj}

The quaternionic qubit can be considered as real subspace of \D4 complex
vector space. Here it is described as real subspace of 4-components
Dirac spinor.

Dirac equation in system of unit $\hbar = 1$, $c = 1$ is \cite{DauIV}:

\begin{equation}
 (i\gm0\pd{t}+i\gm1\pd{x}+i\gm2\pd{y}+i\gm3\pd{z}-m)\varPsi = 0
\label{DirEq}
\end{equation}

Here $\varPsi$ is 4-components complex function and $\gm{i}$ are $4\times4$
complex Dirac matrices are expressed via $2\times2$ Pauli matrices as:
\begin{equation}
\begin{array}{lll}
 & \gm0 = \Mx({0&1\\1&0}) ; \\ \\
 \gm1 = \Mx({0&-\sx\\ \sx&0}) ;& \gm2 = \Mx({0&-\sy\\ \sy&0}) ;&
 \ \gm3 = \Mx({0&-\sz\\ \sz&0})
\end{array}
\label{GmsMat}
\end{equation}
(here $0$ and $1$ are $2\times2$ matrices).

It is also useful to introduce $4\times4\,$ $\gm5$ matrix:
\begin{equation}
 \gm5 = -i \gm0\gm1\gm2\gm3 = \Mx({-1&0\\0&1})
\label{GmVMat}
\end{equation}

The five gamma matrices have following algebraic properties:
\begin{equation}
 \gm{i}\gm{j} = -\gm{j}\gm{i} \,, \ (i \ne j);
 \quad \gm1^2 = \gm2^2 = \gm3^2 = -1; \ \gm0^2 = \gm5^2 = 1
\label{GmAlg}
\end{equation}
The \eq{GmAlg} do not depend on basis. If we choose other representation
with $\varPsi' = \Op{U} \varPsi$ for some unitary operator $\Op{U}$, in
the new basis $\gm{}$ matrices may have other numerical form instead
of \eqs{GmsMat,GmVMat} :
\begin{equation}
\gm{i}' = \Op{U} \gm{i} \Op{U}^{-1} = \Op{U} \gm{i} \Op{U}^\hc
\label{GmsTrans}
\end{equation}
but relations \eq{GmAlg} do not change\footnote{They introduce
Dirac algebra as some abstract object --- \D4 Clifford algebra}.

\medskip

Now it is necessary to find transformations of \D4 spinor $\varPsi$ those
correspond to $SU(2)$ errors of nonrelativistic \D2 Pauli spinor.

As a good candidate here is considered transformations:
$$e_0 \Imat + e_1\gm2\gm3 + e_2\gm3\gm1 + e_3\gm1\gm2$$
where $e_k$ are real numbers, $\Imat$ is $4\times4$ matrix unit and in usual
spinor notation \eq{GmsMat} other three matrices are represented as:
\begin{equation}
  \gm2\gm3 = i\Mx({\sx&0\\0&\sx}) ;
\ \gm3\gm1 = i\Mx({\sy&0\\0&\sy}) ;
\ \gm1\gm2 = i\Mx({\sz&0\\0&\sz})
\label{GmsErMat}
\end{equation}

The representation meets with correspondence principle; 4-spinor can be
expressed as two 2-spinors: $\varPsi = \smx({\xi\\ \eta})$. In so called
{\em standard} representation of Dirac equation are used other two
2-vectors $\varphi = (\xi + \eta)/\sqrt{2}$ and
\EQ{\chi = (\xi - \eta)/\sqrt{2}}. Then for rest particle $\chi = 0$,
{\em i.e.} in nonrelativistic limit: \EQ{v \ll c}, $\chi\approx 0$ can
be omitted and we can work with one Pauli spinor $\varphi$ \cite{DauIV}.

So the matrices \eq{GmsErMat} are only appropriate, because they
do not break condition $\chi = (\xi - \eta)/\sqrt{2} = 0$.

\medskip

The usual Dirac equation \eq{DirEq} cannot be considered as equation
for \D4-real vector $\varPsi$. To make the equation real it is necessary
to find representation with all matrices $i\gm{k}$, $(k = 0,\dots,3)$
are real by some transformation like \eq{GmsTrans}.
Such form of Dirac equation is called by name of Italian
physicist E. Majorana after his work at 1937.

There are many different real representations related via \eq{GmsTrans}
with {\em orthogonal} matrices $\Op{U} \in SO(4) \subset SU(4)$ and here is
used most convenient for particular purpose. An unitary transformation `swaps'
$\gm3 \swap i\gm0$, $\gm1 \swap i\gm5$ (the `$\gm2$' term is used
to change signs of all matrices except $\gm2$) :
\begin{equation}
 \Op{U\!}_M = \gm2\,(\gm5 + i\gm1)(\gm0 + i\gm3)/2 =
\frac{i{+}1}{2}\smx({0&1&i&\,0\,\\-i&0&0&1\\-1&0&0&i\\0&-i&-1&0})
\label{Um}
\end{equation}
\begin{equation}
 \gm0' = -i\gm3;\ \gm1' = i\gm5;\ \gm2'=\gm2;\ \gm3' = i\gm0;
 \ (\mbox{and}\ \gm5' = -i\gm1 \ )
\label{GMm}
\end{equation}

In the new basis expressions for errors \eq{GmsErMat} include
only matrices with all 16 elements are real:
\begin{equation}
  \gm2'\gm3' = \Mx({-i\sy&0\\0&i\sy}) ;
\ \gm3'\gm1' = \Mx({0&-1\\1&0}) ;
\ \gm1'\gm2' = \Mx({0&i\sy\\i\sy&0})
\label{GmsErRMat}
\end{equation}
or
\begin{equation}
  \gm2'\gm3' = \smx({0&\!-1&0&0\\ \!+1&0&0&0\\0&0&0&\!+1\\0&0&\!-1&0}) ;
\ \gm3'\gm1' = \smx({0&0&\!-1&0\\0&0&0&\!-1\\ \!+1&0&0&0\\0&\!+1&0&0}) ;
\ \gm1'\gm2' = \smx({0&0&0&\!+1\\0&0&\!-1&0\\0&\!+1&0&0\\ \!-1&0&0&0})
\label{GmsErRMatExt}
\end{equation}
those correspond to right multiplication on quaternion units represented
as $4\times4$ real matrices, {\em i.e.} for $\q = q_0+q_1\I+q_2\J+q_3\K$:
\begin{equation}
\left.
\begin{array}{rclcl}
 {\q} \cdot \I &=& -q_1+q_0\I+q_3\J-q_2\K
 &\isoeq& -\Op{E}^M_1\ket{\varPsi_{\q}} \\
 {\q} \cdot \J &=& -q_2-q_3\I+q_0\J+q_1\K
 &\isoeq& -\Op{E}^M_2\ket{\varPsi_{\q}}\\
 {\q} \cdot \K &=& -q_3+q_2\I-q_1\J+q_0\K
 &\isoeq& -\Op{E}^M_3\ket{\varPsi_{\q}}
\end{array} \ \right\}
 \ \ \ket{\varPsi_{\q}}\equiv
 {\renewcommand{\arraystretch}{0.75} \Mx({q_0\\q_1\\q_2\\q_3})}
\end{equation}
where $\Op{E}^M_1=-\gm2'\gm3'$, $\Op{E}^M_2=-\gm3'\gm1'$,
$\Op{E}^M_3 = \Op{E}^M_1 \Op{E}^M_2 = \gm1'\gm2'$. Finally:
\begin{equation}
(e_0 + e_1 \Op{E}^M_1 + e_2 \Op{E}^M_2 + e_3 \Op{E}^M_3) \ket{\varPsi_{\q}}
\isoeq \q \, (e_0 - e_1 \I - e_2 \J - e_3 \K) \equiv \q\, \qc{\vc{e}}
\label{EpsiQe}
\end{equation}

\medskip

Because the existence of the `true neutral' Majorana particles are not
proved yet, it is useful to consider relation of the formulae with usual
complex Dirac equation.

In the case, after transformation \eq{Um} of \D4 complex wave function
$\varPsi$ in Dirac equation to new basis, the function:
$$
 \ket{\varPsi'} = \Op{U\!}_M \ket{\varPsi}
$$
may again be complex, but real and imaginary parts of the vector are transformed
separately by errors like \eq{EpsiQe}. If the function $\varPsi'$ prepared as
pure real, it will be real after any such error, {\em i.e.} real subspace of
the complex vector space is {\em invariant} in respect to action of group of
errors \eq{EpsiQe}.

In such a case equation with pure imaginary $\gm{k}'$ \eq{GMm} is considered
as an equivalent form of Dirac equation in other basis with complex-valued:
$$
 \ket{\varPsi'} = \ket{\varPsi_1'} + i\ket{\varPsi_2'}
$$
with two `independent' quaternionic parts $\varPsi_1'$ and $\varPsi_2'$.
It can be also written as:
\begin{equation}
 \Sph^1 \times \Sph^3 \:\subset\: \C\ox_\R\H \:\iso\: \H\oplus\H
\label{SHqub}
\end{equation}
where $\Sph^3$ and $\H$ are spaces related with $\h$-qubits and
$\Sph^1$ is phase. The complexification \eq{SHqub} of $\h$-qubit is represented
as simple direct product ({\em i.e.} trivial bundle) in comparison with
nontrivial Hopf fibration of usual $\C^2$-qubit.

\section{Conclusion}

As not very formal answer to the question, why 3 is enough for classical case,
but is not enough in quantum one, may be used suggestion that \D2 space $\Sph^2$
of one qubit is `too small' in comparison with \D3 space of all possible errors.
In classical case we have only one possible error --- flip of a bit. In case
with $\R$-qubit it is \D1 space ($\Sph^1$) with \D1 space of errors and in
$\H$ case it is \D3 space ($\Sph^3$) with \D3 space of errors.

The idea is also related with initial Shor's 9-qubit code \cite{ShEC}, because
the code can be considered as two-steps process: first, we {\em pre\/}encode
qubit to 3 qubits:
$$
\ket{0} \stackrel{pre}{\dasharrow} \ket{B_3^+}=(\ket{000}+\ket{111})/\sqrt{2},
\ \ \ket{1} \stackrel{pre}{\dasharrow} \ket{B_3^-}=(\ket{000}-\ket{111})/\sqrt{2}
$$
The code $B^\pm_3$ belongs to \D8 space, has 7 different\footnote{Here
$7 = 3 \times 3 - 2$ (all three phase errors with different qubits act in
the same way)} kinds of errors and second step is repeating:
$$
\ket{0} \to \ket{B_3^+ B_3^+ B_3^+},
\quad \ket{1} \to \ket{B_3^- B_3^- B_3^-}
$$

Then the last example with Dirac spinors can be considered as an analogue
of Shor idea, but with {\em pre\/}encoding due to additional physical degrees
of freedom of relativistic particle in some subspace invariant with respect
to \D3 group of `slack', nonrelativistic errors. Here is only question, do
the {\em pre\/}encoding and error correction operator $\Op{U}_{ec}$ physically
possible --- they are unitary, but not all unitary operation with
relativistic particles would be performed under realistic conditions.

\end{document}